\documentclass[onecolumn,showpacs,preprintnumbers,amssymb]{revtex4}
\def\ii{{\rm i}}
\usepackage{latexsym}
\usepackage{graphicx,epsf, epsfig, amssymb}
\usepackage{bm}
\usepackage{longtable}

\def\be{\begin{equation}}
\def\ee{\end{equation}}
\def\beq{\begin{eqnarray}}
\def\eeq{\end{eqnarray}}

\begin{document}

\title{Perturbations of Schwarzschild black holes in\\ Dynamical Chern-Simons modified gravity}

\author{Vitor Cardoso} \email{vitor.cardoso@ist.utl.pt} \affiliation{Centro Multidisciplinar de
  Astrof\'{\i}sica - CENTRA, Dept. de F\'{\i}sica, Instituto Superior T\'ecnico, Av. Rovisco Pais 1,
  1049-001 Lisboa, Portugal \&\\ Department of Physics and Astronomy, The University of Mississippi,
  University, MS 38677-1848, USA}

\author{Leonardo Gualtieri} \email{Leonardo.Gualtieri@roma1.infn.it} \affiliation{Dipartimento di
  Fisica, Universit\`a di Roma ``Sapienza'' \& Sezione INFN Roma1, P.A. Moro 5, 00185, Roma, Italy}

\begin{abstract} 
Dynamical Chern-Simons (DCS) modified gravity is an attractive, yet relatively unexplored,
candidate to an alternative theory of gravity. The DCS correction couples a dynamical scalar field
to the gravitational field. In this framework, we analyze the perturbation formalism and stability
properties of spherically symmetric black holes. Assuming that no background scalar field is
present, gravitational perturbations with polar and axial parities decouple. We find no effect of
the Chern-Simons coupling on the polar sector, while axial perturbations couple to the Chern-Simons
scalar field. The axial sector can develop strong instabilities if the coupling parameter $\beta$,
associated to the dynamical coupling of the scalar field, is small enough; this yields a
constraint on $\beta$ which is much stronger than the constraints previously known in the
literature.
\end{abstract}

\pacs{04.50.Kd,
04.25.-g,
04.25.Nk,04.30.-w
}
\maketitle

\section{Introduction}\label{intro}
Interferometric and resonant-bar gravitational-wave detectors are now working at or near design
sensitivity. It is expected that instruments such as LIGO or any of its advanced versions
\cite{:2007kva} will soon make the first direct detection of gravitational waves on Earth. The
detection and measurement of gravitational waves from compact, massive astrophysical bodies opens a
new window into the universe, and it also opens up the exciting possibility of testing Einstein's
field equations in an unprecedented way. From measurements of the inspiralling phase of black hole
or neutron star binaries, one can test General Relativity's prediction for the waveform and among
others, bound the mass of the graviton \cite{Will:2004xi,Berti:2004bd,Stavridis:2009mb}.
Gravitational-wave observations of the ringdown phase of the final black hole can lead to tests of
the no-hair theorem in General Relativity \cite{Berti:2005ys,Berti:2009kk}. Taken together, these
measurements also allow for tests of Hawking's area theorem for black holes
\cite{Sathyaprakash:2009xs}.

What if there are corrections to the field equations, and do they leave a measurable imprint on
gravitational waves? Unfortunately, this important question depends on several unknowns. It depends
on the form of the corrections to the field equations, on the black hole solutions to these modified
equations and on which specific process is generating the gravitational wave signal. One of such
theories is known as Einstein-Dilatonic-Gauss-Bonnet, a simple example of one-loop corrected
four-dimensional effective theory of the heterotic superstrings at low energies \cite{Moura:2006pz}.
The possibility of astrophysical tests of this theory were studied in Ref. \cite{Pani:2009wy}.

Another promising extension of general relativity is Chern-Simons (CS) gravity \cite{Lue:1998mq,Jackiw:2003pm,Alexander:2009tp}, in which the Einstein-Hilbert action is modified by adding a parity-violating Chern-Simons term, which couples to gravity via a scalar field. This correction arises in many contexts. Such a term could help to explain several problems of cosmology, from inflation (as discussed by Weinberg \cite{Weinberg:2008mc}) to baryon asymmetry \cite{GarciaBellido:2003wd,Alexander:2004xd,Alexander:2004us}. In most of the moduli space of string theory, a CS correction is required to preserve unitarity; furthermore, duality symmetries induce a CS term in all string theories with a Ramond-Ramond scalar \cite{Polchinski:1998rr}. In loop quantum gravity, it is required to ensure gauge invariance of the Ashtekar variables \cite{Ashtekar:1988sw} and it also arises naturally if the Barbero-Immirzi parameter is promoted to a field \cite{Taveras:2008yf,Mercuri:2009zt}.

An interesting feature of CS modified gravity is that it has a characteristic observational
signature, which could allow one to discriminate an effect of this theory from other phenomena. Indeed,
the CS term violates parity, and thus it mainly affects the axial-parity component of the
gravitational field. For instance, it yields amplitude birefringence of gravitational waves (the two
polarizations travel with the same speed, but one is enhanced, while the other is suppressed); on the
other hand, the Schwarzschild solution is unaffected by CS modified gravity, and then the solar
system tests of general relativity do not put strict bounds on the magnitude of this correction.
Another possible signature of CS modified gravity may be found in the polarization of primordial gravitational waves \cite{Satoh:2007gn}.

Most of the literature on CS modified gravity refers to its non-dynamical formulation, in which the
scalar field is a prescribed function. Usually, the so-called canonical prescription $\theta\propto
t$ is chosen \cite{Alexander:2007kv,Alexander:2007zg,Alexander:2007vt}. An explicit prescription for
the scalar field is necessarily {\it a priori}, and furthermore it breaks gauge invariance (see the
discussion in \cite{Yunes:2009hc}). For these reasons, we prefer to consider the dynamical
Chern-Simons (DCS) modified gravity \cite{Smith:2007jm}, where the scalar field is treated as a
dynamical field. Note that CS gravity and DCS gravity are inequivalent and independent theories:
although the CS gravity action can be obtained as a certain limit of the DCS gravity action, the
solutions of CS gravity cannot be obtained from the solutions of DCS gravity \cite{Yunes:2009hc}. We
also remark that CS modified gravity, with the canonical prescription, is retrieved in the case of
weak gravitational field, if one only keeps the leading order perturbations of Minkowski spacetime
\cite{Yunes:2008ua}; therefore, CS gravity is appropriate to study, for instance, deviations from
general relativity in solar system astrophysical processes, or in the motion of binary pulsars far
from coalescence, but DCS gravity is required to describe deviation from general relativity in the
strong field regime, like in black hole (BH) perturbation theory or in the coalescence of binary
systems.

In this paper we study, in the context of DCS gravity, gravitational perturbations of a spherically
symmetric black hole. Perturbations of black holes are interesting for several reasons: since black
holes populate the universe in large number, assessing their stability in a given theory is
tantamount to testing the theory: if the black hole solution in that specific theory is unstable,
they would not be seen. Second because a variety of processes taking place in the vicinities of
black holes will be observed by gravitational-wave detectors. The two most important are
extreme-mass-ratio-inspirals \cite{AmaroSeoane:2007aw}, and quasi-normal ringing
\cite{Berti:2009kk}. The first consists, for instance, on a small star orbiting around a
supermassive black hole. Such a process can be modeled as a test particle inducing perturbations on
a black hole background, and can be tackled with perturbation tools. The second is a universal
signal: all or almost all events involving black holes produce a gravitational wave signal which at
late times consists of a superposition of the characteristic modes of the black hole, the
quasi-normal modes (QNMs). Such a signal will be seen by present or future gravitational wave
detectors.
  
BH perturbations have been studied, in the context of CS gravity, in Ref. \cite{Yunes:2007ss}, where
it was found that polar-parity and axial-parity gravitational perturbations are coupled, and the
equations do not allow for generic black hole oscillations. BH perturbations in DCS gravity were
also briefly discussed in that work, in the framework of two-parameter perturbation theory: the two
parameters were $\epsilon$, describing the magnitude of the gravitational perturbation, and $\tau$,
describing the magnitude of the CS scalar field. The authors of \cite{Yunes:2007ss} found that, if
the scalar field has both a background (spherically symmetric) component of order $O(\tau)$ and an
oscillating component of order $O(\tau\epsilon)$, polar and axial perturbations are coupled, and the
equations are extremely involved. Here, we consider the case in which the background scalar field
vanishes, i.e. the scalar field is only generated by gravitational perturbations. Under this
condition, we find that polar and axial gravitational perturbations decouple, and only axial parity
gravitational perturbations are affected by the CS scalar field. Furthermore we find that, under
this assumption, gravitational perturbations and the Chern-Simons scalar field are described by a
simple set of equations. Numerical integration of these equations is not an easy task, due to the
well-known asymptotic divergence which prevented for many years the numerical computation of
quasi-normal modes of the Schwarzschild BH \cite{Chandrasekhar:1975zz,Nollert:1999ji}.

We find some evidence that in the limit $\beta\rightarrow\infty$, where $\beta$ is the coefficient
in front of the kinetic CS term of the action (see Eq. (\ref{action}) and discussion below), the BH
does not admit QNMs. Furthermore we find that for $0<\beta\,M^4\lesssim2\pi$, there is at least one
(strongly) unstable mode, and the Schwarzschild DCS solution is then unstable. This seems to impose
a strong constraint on the theory, if it is to be compatible with the existence of astrophysical
black holes, i.e., $\beta\gtrsim10^{-2}$ km$^{-4}$. Previous bounds were much weaker: in
\cite{Yunes:2009hc} it was found that observational data from the extreme double pulsar system PSR
J0737-3039 imply $\beta\gtrsim10^{-15}$ km$^{-4}$.

Strictly speaking, our results apply only to non-rotating BHs, since rotating BH have a non-vanishing background scalar field \cite{Yunes:2009hc,Konno:2009kg} and are more complex to handle. Astrophysical BHs are in general rotating (perhaps even rapidly rotating), but it is reasonable to expect that the range of $\beta$ incompatible with the existence of non-rotating BH is also incompatible with (slowly) rotating BH. Therefore, we take the bound $\beta\gtrsim10^{-2}$ km$^{-4}$ as a (strong) indication of the values we may expect for the parameter $\beta$.
Finally, we also want to stress that the bound $\beta\gtrsim10^{-2}$ km$^{-4}$ does not at all rule out the possibility of an observational signature from DCS gravity.  Indeed, we restrict the range of $\beta$, but we do not restrict neither the range of $\alpha$ (i.e., with the $\alpha=1$ normalization, the amplitude of the scalar field), nor the time derivative of the scalar field amplitude. For instance, in \cite{Yunes:2008ua} it was found that observational data from the double pulsar J0737-3039 imply a bound on $\dot\vartheta$, in the context of CS gravity.

The paper is organized as follows. In Section \ref{DCS} we briefly review DCS modified gravity. In
Section \ref{pert} we derive the equations for perturbations of a spherically symmetric black hole
in DCS modified gravity. In Section \ref{seceq} we discuss the perturbation equations we have
derived, finding some of their solutions. In Section \ref{concl} we draw our conclusions.

\section{DCS gravity}\label{DCS}

Following the notation of \cite{Yunes:2009hc}, the action of dynamical Chern-Simons modified gravity
is
\begin{equation}
S=\kappa\int d^4x\sqrt{-g}R+\frac{\alpha}{4}\int d^4x\sqrt{-g}
\vartheta\,^*RR-\frac{\beta}{2}\int d^4x\sqrt{-g}\left[
g^{ab}\nabla_a\vartheta\nabla_b\vartheta+V(\vartheta)\right]+S_{mat}\,.\label{action}
\end{equation}
Note that there are two parameters $\alpha,\beta$, but one of them can be eliminated by choosing the
normalization of the scalar field. For instance, in \cite{Smith:2007jm} the normalization of
$\vartheta$ is chosen such that $\beta=1$, and $\alpha$ is the only parameter describing the
coupling between the scalar field and the gravitational field. We normalize instead the scalar field
by imposing $\alpha=1$, and use geometrical units $c=G=1$ so that $\kappa=\frac{1}{16\pi}$. Then,
\begin{equation} 
[S]=l^2\,,~~~~~[\vartheta]=l^2\,,~~~~[\beta]=l^{-4}\,. 
\end{equation} 
Furthermore, we neglect $V(\vartheta)$, and consider the vacuum solutions ($S_{mat}=0$); therefore,
the equations of motion are
\begin{eqnarray}
  R_{ab}&=&-16\pi C_{ab}+8\pi\beta\vartheta_{,a}\vartheta_{,b}\label{eqE1}\\
\Box\vartheta&=&-\frac{1}{4\beta}\,^*RR \label{eqE2}
\end{eqnarray}
where
\begin{eqnarray}
C^{ab}&=&\vartheta_{;a}\epsilon^{cde(a}\nabla_eR^{b)}_{~~d}
+\vartheta_{;dc}\,^*R^{d(ab)c}\\
^*RR&=&\frac{1}{2}R_{abcd}\epsilon^{abef}R^{cd}_{~~ef}\,.
\end{eqnarray}
Equations (\ref{eqE1}), (\ref{eqE2}) acquire a particularly simple form in the spherically symmetric case \cite{Jackiw:2003pm,Yunes:2007ss}, yielding (up to $O(\vartheta^2)$) the Schwarzschild solution; indeed, in the Schwarzschild spacetime $^*RR=0$ and, assuming that the scalar field is also spherically symmetric $\vartheta=\vartheta(t,r)$, $C^{ab}=0$. Then, if we neglect the CS stress-energy tensor (quadratic in the scalar field), Eqns. (\ref{eqE1}), (\ref{eqE2}) are satisfied by the Schwarzschild solution.

We mention that in \cite{Yunes:2009hc,Konno:2009kg} the solution for slowly rotating black holes has
been found in DCS modified gravity; the corrections from the general relativistic solutions are of
the order $\alpha/\beta$, $\alpha^2/(\beta\kappa)$, i.e., with our normalizations, of the order
$\beta^{-1}$. Furthermore, an observational constraint for $\beta^{-1}$ has been derived in
\cite{Yunes:2009hc}, from frame dragging effects in the extreme double pulsar system PSR J0737-3039
A/B:
\begin{equation}
\beta^{-1}\lesssim 10^{15}~{\rm km}^4\,.\label{ysconstr}
\end{equation}
\section{Perturbations of a Schwarzschild background}\label{pert}
We now consider perturbations of the spacetime geometry away from a Schwarzschild background. We set
$g^{(0)}_{\mu\nu}$ to be the Schwarzschild metric and choose the Regge-Wheeler gauge for
perturbations:
\begin{eqnarray}
g_{\mu\nu}&=&g^{(0)}_{\mu\nu}+h_{\mu\nu}\,,\nonumber\\
g^{(0)}_{\mu\nu}&=&{\rm diag}(-f,f^{-1},r^2,r^2\sin^2\theta)
~~~~~\left(~f(r)\equiv1-2M/r~\right)\,,\nonumber\\
h_{\mu\nu}&=&\left(\begin{array}{cc|cc}
H_0^{lm}Y^{lm}&H_1^{lm}Y^{lm}&h_0^{lm}S_\theta^{lm}
&h_0^{lm}S_\phi^{lm}\\
H_1^{lm}Y^{lm}&H_2^{lm}Y^{lm}&h_1^{lm}S_\theta^{lm}
&h_1^{lm}S_\phi^{lm}\\\hline
h_0^{lm}S_\theta^{lm}&h_1^{lm}S_\theta^{lm}&r^2K^{lm}Y^{lm}&0\\
h_0^{lm}S_\phi^{lm}&h_1^{lm}S_\phi^{lm}&0&r^2K^{lm}\sin^2\theta Y^{lm}\\
\end{array}\right)e^{-\ii\omega t}\,,\label{expmetric}
\end{eqnarray}
where $Y^{lm}$ are the scalar spherical harmonics and
\begin{equation}
(S_\theta^{lm},S_\phi^{lm})\equiv\left(-\frac{1}{\sin\theta}
Y^{lm}_{,\phi},\sin\theta Y^{lm}_{,\theta}\right)\,.
\end{equation}
Here, $(H_0,H_1,H_2,K)^{lm}$ are (functions of $r$) describing the polar parity metric
perturbations, $(h_0,h_1)^{lm}$ describe the axial parity metric perturbations.

We now make one simplifying, self-consistent assumption, namely that the Chern-Simons scalar field
is of the order of $O(h)$. In other words, the background scalar field $\vartheta^{(0)}$, the solution of the
homogeneous equation $\Box\vartheta^{(0)}=0$, is vanishing: $\vartheta^{(0)}\equiv0$. The only
scalar field present is induced by the perturbations. In this respect the case we are considering is
different from that considered in Section VI of Ref. \cite{Yunes:2007ss}. Under this assumption, the
harmonic expansion of the scalar field $\vartheta$ is
\begin{equation}
\vartheta=\frac{\Theta}{r}Y^{lm}e^{-\ii\omega t}\,.\label{expscalar}
\end{equation}
From here onwards, we will drop the $^{lm}$ superscripts.

Replacing expansions (\ref{expmetric}), (\ref{expscalar}) into Eqs. (\ref{eqE1}), (\ref{eqE2}) and
neglecting terms quadratic in the metric perturbations or in the scalar field, we find that
perturbations with different parities decouple: polar parity metric perturbations are unaffected by
the scalar field. These will not be discussed any further here, since the stability properties and
QNMs of these solutions are very well known (see \cite{Chandrasekhar:1985kt} and references therein;
for recent reviews see \cite{Ferrari:2007dd,Berti:2009kk}). On the other hand, the axial parity
metric perturbations are coupled with the scalar field. Indeed, equations (\ref{eqE1}) yield
\begin{eqnarray}
E_1&\equiv&r^3\left(-4M+l(l+1)r\right)h_0-rf\left(2ir^4\omega \,h_1+192\pi\,M\Theta+ir^5\omega h_1'-96\pi\,Mr\Theta'+r^5h_0''\right)=0\,,\label{eqEi1}\\
E_2&\equiv& -i\omega r^3\left(2h_0-ir\omega\,h_1-rh_0'\right)+r^2f\left(l^2+l-2\right)h_1-96\pi\,i\,M\omega\Theta=0\,,\label{eqEi2}\\
E_3&\equiv&ir^3\omega \,h_0+rf \left(+2M h_1+r^2f h_1'\right)=0\,.\label{eqEi3}
\end{eqnarray}
We note that these three equations are not all independent, it is easy to show that 
\begin{equation}
-\frac{fr^4}{i\omega}\left (E_2/r^2\right)'-E_3+\frac{(l-1)(l+2)\,r}{i\omega} E_1=0\,.
\end{equation}
Combining (\ref{eqEi1}), (\ref{eqEi2}) and (\ref{eqEi3}) and defining the Regge-Wheeler master
function $Q(r)$ by
\begin{equation}
h_1=f^{-1}r\,Q\,.
\end{equation}
and
\begin{equation}
r_*\equiv r+2M\ln\left(r/2M-1\right)
\end{equation}
we finally get
\begin{eqnarray}
\frac{d^2}{dr_*^2}Q+\left[\omega^2-f\left(\frac{l(l+1)}{r^2}-\frac{6M}{r^3}\right)\right]Q&=&
-\frac{96\pi\,iMf\omega}{r^5} \Theta\,,\label{eqq}\\
\frac{d^2}{dr_*^2}\Theta+\left[\omega^2-f\left(\frac{l(l+1)}{r^2}
\left(1-\frac{576\pi M^2}{r^6\beta}\right)
+\frac{2M}{r^3}\right)\right]\Theta&=&-f\frac{(l+2)!}{(l-2)!}\frac{6\ii M}{\omega r^5\beta}
Q\,.\label{eqpsi}
\end{eqnarray}
Equations (\ref{eqq}) and (\ref{eqpsi}) form a system of coupled second order differential equations
for the perturbations $Q^{lm}$, $\Theta^{lm}$ (with dimensions $[Q^{lm}]=l^0$, $[\Theta^{lm}]=l^3$),
from which one can completely characterize the axial parity metric perturbations and the scalar
field.
\section{Integration of the perturbation equations}\label{seceq}
Despite their apparent simplicity, numerical integration of the perturbation equations (\ref{eqq}),
(\ref{eqpsi}) is not an easy task. We first note that, as $r_*\rightarrow\pm\infty$ (i.e.
$r\rightarrow r_H\equiv2M,r\rightarrow+\infty$), they reduce to the simple wave equations
\begin{equation}
\left(\frac{d^2}{dr_*^2}+\omega^2\right)\Theta^{lm}=
\left(\frac{d^2}{dr_*^2}+\omega^2\right)Q^{lm}=0\,;\label{limitpm}
\end{equation} 
then, every solution $(Q,\Theta)$ of the perturbation equations has the asymptotic form,
as $r_*\rightarrow\pm\infty$,
\begin{eqnarray}
Q&=&A_{H,\infty}^{\rm out}e^{\ii\omega r_*}+
A_{H,\infty}^{\rm in}e^{-\ii\omega r_*}\,,\nonumber\\
\Theta&=&B_{H,\infty}^{\rm out}e^{\ii\omega r_*}+
B_{H,\infty}^{\rm in}e^{-\ii\omega r_*}\,,\label{behaviorboundaries}
\end{eqnarray}
where $\omega,A_{H,\infty}^{\rm out,in},B_{H,\infty}^{\rm out,in}$ are complex numbers.  A QNM is a
solution of the perturbation equation which satisfies the Sommerfeld boundary conditions
\begin{equation}
A_H^{\rm out}=B_H^{\rm out}=A_\infty^{\rm in}=B_\infty^{\rm in}=0\,,\label{bc}
\end{equation}
i.e., no radiation outgoing from the horizon, no radiation ingoing from infinity. 

By fixing the normalization $B_H^{\rm in}=1$, and defining $A_H^{\rm in}=A_0\in~ !\!\!\!C$, we have
a unique solution of equations (\ref{eqq}) and (\ref{eqpsi}); thus we end with two complex
conditions ($A^{\rm in}_\infty=B_\infty^{\rm in}=0$) to satisfy, and two complex numbers to
determine ($A_0$ and $\omega$). This naive counting of degrees of freedom tells us that, as in the
case of the Schwarzschild BH, the dimension of the space of solutions is zero, allowing for a discrete
set of complex values for $\omega$: the quasinormal modes of the black hole. The difference
from the Schwarzschild case is that here there are two coupled differential equations, instead of a
single Schroedinger-like equation.

We are not aware of any discussion in the literature regarding the QNMs of coupled systems of ODEs
such as the one we have here. The main difficulty in integrating equations (\ref{eqq}) and
(\ref{eqpsi}) is known: for stable spacetimes $Im(\omega)<0$ and the correct behavior at (say)
infinity is $e^{i\omega r_*}$, which is exponentially dominant over the unwanted $e^{-\ii\omega
  r_*}$. In the case of the Schwarzschild BH, many routes have been explored to circumvent the divergence
problem discussed above; the main successful approaches are: a reformulation of the equation as a
recurrence relation, which is then expressed as a continued fraction \cite{Leaver:1985ax}; a WKB
approximation of the potential \cite{Iyer:1986np}; an analytic continuation to the complex plane of
the radial coordinate \cite{Andersson:1992a}. It seems quite difficult to extend any of these
approaches to a system of two coupled equations like (\ref{eqq}) and (\ref{eqpsi}). For instance, if
one reformulates our perturbation equations in terms of recurrence series (see Appendix
\ref{recurrence}), one finds two coupled series, with ten terms at each order, which do not seem to
be expressible in terms of continued fractions.

Therefore, we were not able to perform a full, numerically accurate search for the QNMs of the
coupled system (\ref{eqq}) and (\ref{eqpsi}). We studied only two limiting regimes: $\beta M^4 \gg1$
and $\beta M^4\lesssim1$. Before proceeding to the analysis of these two cases, we mention that,
since $Q,\Theta$ exponentially diverge in the $r_*\rightarrow\pm\infty$ limits, it is not at all
obvious that equations (\ref{eqq}), (\ref{eqpsi}) reduce to the form (\ref{limitpm}) and then that
their solutions, in these limits, have the form (\ref{behaviorboundaries}), as we have assumed at
the beginning of this section. Actually, Eq.(\ref{behaviorboundaries}) can be proved without
assuming (\ref{limitpm}), as we show in Appendix \ref{app:asympt}.
\subsection{Very large $\beta$}
In the case $\beta M^4\gg1$ Eqns. (\ref{eqq}), (\ref{eqpsi}) reduce to
\begin{eqnarray}
&&\frac{d^2}{dr_*^2}Q^{lm}+\left[\omega^2-f\left(\frac{l(l+1)}{r^2}
-\frac{6M}{r^3}\right)\right]Q^{lm}=-f\frac{96\pi\ii\omega M}{r^5}
\Theta^{lm}\label{qlargebeta}\\
&&\frac{d^2}{dr_*^2}\Theta^{lm}+\left[\omega^2-f\left(\frac{l(l+1)}{r^2}
+\frac{2M}{r^3}\right)\right]\Theta^{lm}=0\,.\label{psilargebeta}
\end{eqnarray}
Thus, in this limit, the eigenvalue problem for $\Theta$ is homogeneous, and exactly equivalent to the Schwarzschild case. The solutions of (\ref{psilargebeta}), $\omega^{QNM},\,\Theta^{QNM}$, are well-known and can be computed accurately using the continued fraction approach \cite{Leaver:1985ax,Berti:2009kk}. Then, we are left with equation (\ref{qlargebeta}), which can be considered as a single differential equation with source. One can build the solution to the inhomogeneous problem (\ref{qlargebeta}) by using the Green function method:
\begin{eqnarray}
Q=\frac{1}{{\cal W}} \left[ Q_{\infty} \int _{2M}^{r}Q_H S dr +
Q_H\int _{r}^{\infty}Q_{\infty} S dr \right]\,,\label{inhomo}
\end{eqnarray}
where $Q_H,\,Q_{\infty}$ are two independent solutions of the homogeneous version of (\ref{qlargebeta}) (evaluated with the $\omega=\omega^{QNM}$ eigenvalue of the scalar equation), such that $Q_{\infty}\sim e^{i\omega r_*}$ at infinity and $Q_H \sim e^{-i\omega r_*}$ close to the horizon; the quantity ${\cal W}$ is the wronskian between $Q_{H},\,Q_{\infty}$, and $S\equiv -\Theta^{QNM}96\pi\ii\omega M/r^5$.

Solution (\ref{inhomo}) can be seen to satisfy the required boundary conditions and indeed is the only solution satisfying such boundary conditions. Unfortunately, the integrals in (\ref{inhomo}) are not well defined: the wavefunctions $Q_H,\,Q_{\infty}$ and $S$ diverge exponentially at both the horizon and spatial infinity. This seems to indicate that this solution does not exist, i.e. that DCS Schwarzschild BH do not admit QNM in which both the gravitational perturbations and the scalar field are excited. Only trivial QNM solutions, in which $\Theta\equiv0$ and the gravitational field oscillates at the frequencies of ordinary Schwarzschild QNM, seem to be allowed in this limit.
\subsection{Instability regime}
\begin{center}
\begin{figure}[ht]
\begin{tabular}{cc}
\epsfig{file=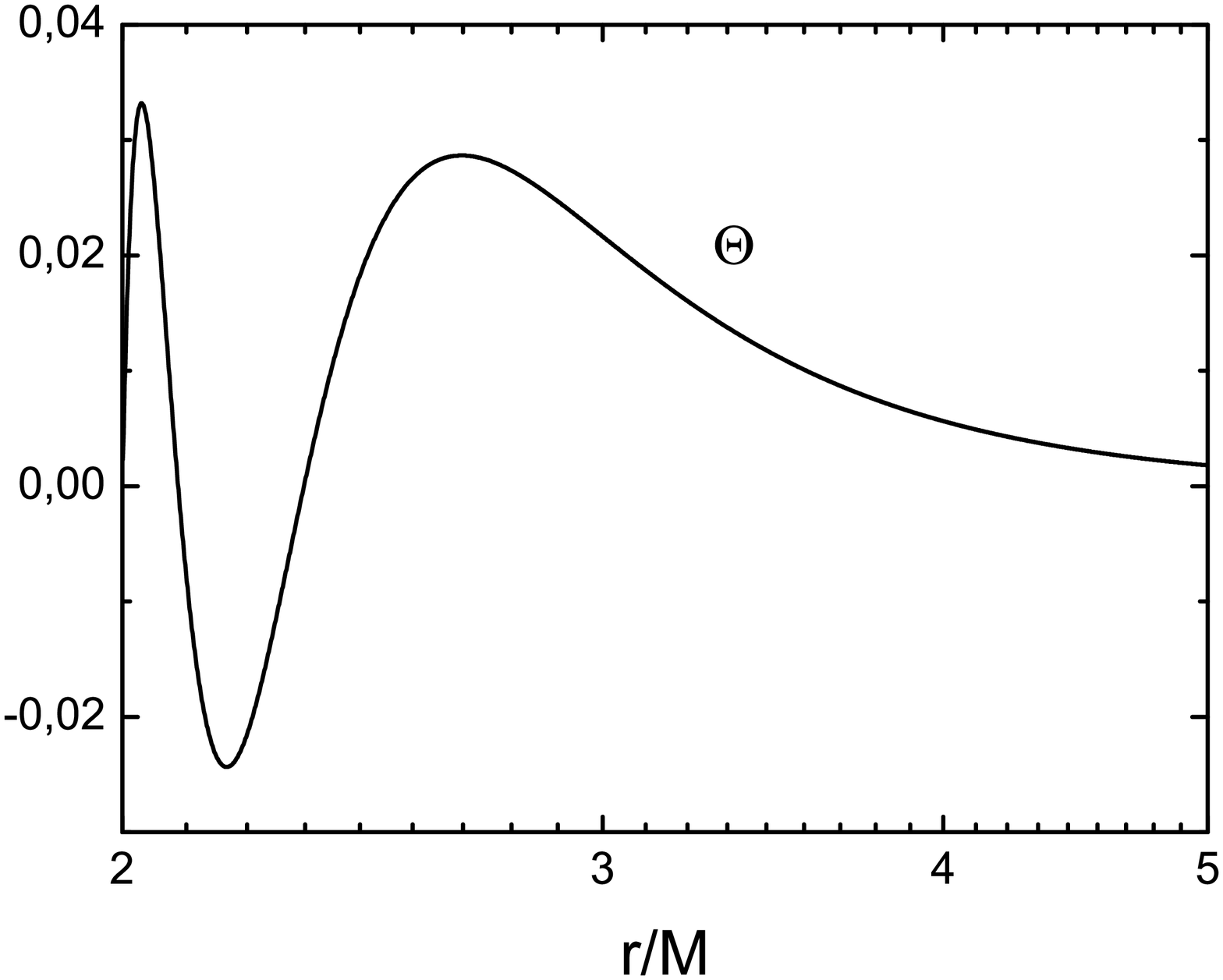,width=250pt,angle=0}&
\epsfig{file=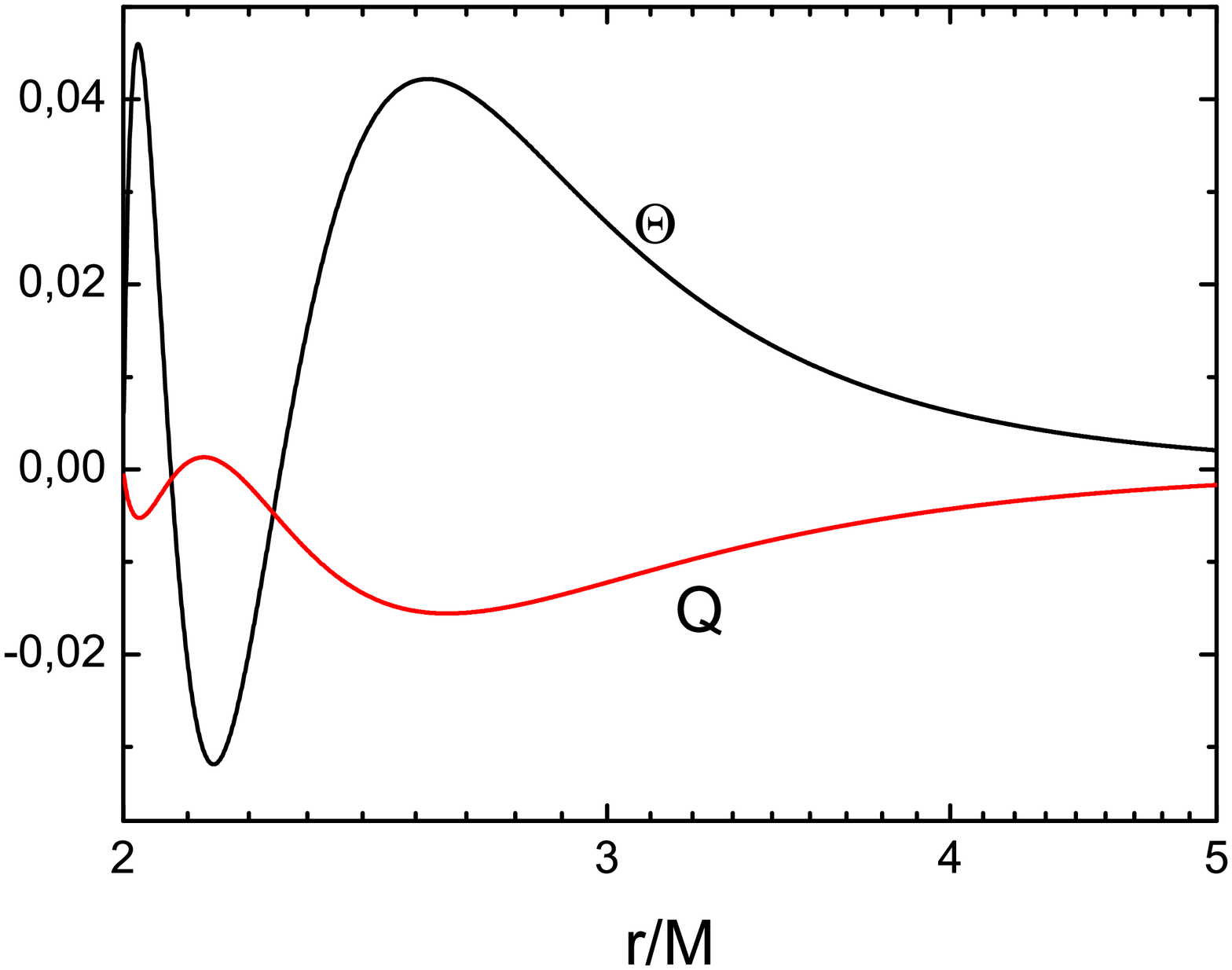,width=250pt,angle=0}
\end{tabular}
\caption{Left: Scalar wavefunction $\Theta$ for the unstable mode corresponding to $M^4\beta=1$ and
  $M\omega=0.51197595\ii$, assuming no gravitational coupling, i.e., this is an unstable mode of the homogeneous version of eq. (\ref{homsc}). Right: Scalar wavefunction $\Theta$ and Regge-Wheeler function $Q$ for the unstable mode of the full coupled system, corresponding to $M^4\beta=1$ and
  $M\omega=0.4280636\ii$, $A_H^{\rm in}=-0.1069141$.\label{fig:1} }
\label{fig:smodes}
\end{figure}
\end{center}

The opposite regime concerns very small $\beta$. As might be anticipated from the previous
discussion, the behavior of the solutions of the coupled system are intimately connected with the
behavior of the homogeneous problem. Now, in the case of small $\beta$, the equation for the scalar
field is
\begin{equation}
\frac{d^2}{dr_*^2}\Theta^{lm}+\left[\omega^2-V(r)\right]\Theta^{lm}=-f\frac{(l+2)!}{(l-2)!}\frac{6\ii M}{\omega r^5\beta}
Q^{lm}\,,\label{homsc}
\end{equation}
with $V=f\left(l(l+1)/r^2\left(1-576\pi M^2\beta^{-1}/r^6\right) +2M/r^3\right)$. A study of the
homogeneous equation (i.e., we set $Q=0$ in the above), shows that unstable modes are possible since
the effective potential is negative in some regions. Since the potential is bounded in
$-\infty<r_*<+\infty$ and it vanishes at both extrema, a sufficient condition for existence of a
bound state of negative energy is \cite{buell,Dotti:2004sh}
\begin{equation}
\int_{-\infty}^{+\infty}V(r_*)dr_*=\frac{7\beta(2l^2+2l+1)M^4-18l(l+1)\pi}{28\beta\,M^5}<0\,.
\end{equation}
This condition is satisfied whenever 
\begin{equation}
\beta<\frac{18 l(l+1)\pi}{7(2l^2+2l+1)M^4}\,,
\end{equation}
which yields $\beta<108\pi/(91M^4),\,216\pi/(175M^4)$ for $l=2,3$ respectively. An example of an
unstable $l=2$ mode of the homogeneous equation is shown in Figure \ref{fig:1}, for $\beta=1$. This
does not prove the existence of unstable modes for the coupled system, but it is a strong argument
in its favor.

Numerical integration of the coupled system (\ref{eqq}), (\ref{eqpsi}) shows that unstable modes, indeed, do exist. In our search, we set $B_H^{\rm in}=1$, and look for solutions with $\omega$ purely imaginary; in this case, the boundary conditions (\ref{bc}) imply that the functions $Q,\Theta$ vanish at $r\rightarrow\infty$. 
A typical unstable mode of the coupled system is shown in Figure \ref{fig:1}, for $\beta=1$. The
overall qualitative and quantitative behavior of the scalar field $\Theta$ is in agreement with the
analysis of the homogeneous scalar wave equation. The instability timescale increases (i.e. ${\rm
  Im}\,\omega$ decreases) when $\beta$ increases. For instance, $M\omega\sim 0.428\ii,0.263\ii$ for $M^4\beta=1,2$ respectively. This was to be expected from the analysis of the homogeneous scalar equation. We were not able to find unstable modes for $\beta M^4$ larger than $\sim2\pi$. This is again in agreement with the heuristic arguments above, though a more systematic approach to investigate this issue is necessary.

We have then shown that spherically symmetric BH in DCS modified gravity admit unstable modes for
$\beta M^4\lesssim2\pi$; therefore, since astrophysical black holes exist with masses
$M_{astro}\gtrsim 3M_\odot\simeq 5$ km, we have the constraint
\begin{equation}
\beta\gtrsim\frac{2\pi}{M_{astro}^4}\gtrsim10^{-2}{\rm km}^{-4}\,,\label{ourbound}
\end{equation} 
much stronger than (\ref{ysconstr}). The present analysis does not directly apply to astrophysical BHs, since these are rotating in general.
Rotating black holes in this theory require a non-vanishing background scalar field \cite{Yunes:2009hc,Konno:2009kg} and therefore our entire analysis needs to be modified. On the other hand, it is hard to imagine that rotation can drastically alter the instability regime.
We thus believe that the bound (\ref{ourbound}) should be considered as a strong indication on the values that we may expect for $\beta$
in the general case.

We remark that the normalization choice $\alpha=1$ does not really affect our results. If we do not fix the parameter $\alpha$, some terms in equations (\ref{eqq}), (\ref{eqpsi}) get multiplied by $\alpha$ or $\alpha^2$, and the bound (\ref{ourbound}) becomes
\begin{equation}
\frac{\beta}{\alpha^2}\gtrsim\frac{2\pi}{M_{astro}^4}\gtrsim10^{-2}{\rm km}^{-4}\,.\label{ourbound2}
\end{equation}
Note that the quantity $\alpha^2/(\beta\kappa)$ is what is called $\xi$ in \cite{Yunes:2009hc}, where the bound (\ref{ysconstr}) is derived. 

\section{Conclusions}\label{concl}
We have studied perturbations of black holes in the context of dynamical Chern-Simons theory. Under
the assumption that the scalar field is of the order of the gravitational perturbation, we found
that axial and polar parity gravitational perturbations are decoupled, and only axial parity
perturbations are coupled with the scalar field. The equations describing these perturbations are
fairly simple, but their numerical integration is tricky: our attempts to generalize to DCS gravity
either the recurrence relation approach of \cite{Leaver:1985ax,Nollert:1993zz}, or the Riccati
equation approach of \cite{Chandrasekhar:1975zz}, have been unsuccessful. Still, we have been able
to extract some interesting information from the perturbation equations.

We found that black holes in this theory can develop strong instabilities. The constraint on the coupling parameters necessary to avoid the instability is stringent. We obtain $\beta M^4 >2\pi$ for the spacetime to be stable, which translates to $\beta\gtrsim10^{-2}$ km$^{-4}$. Thus, the observation of stellar-mass black holes imposes a constraint on the coupling parameter which is about $10^{13}$ times more stringent than bounds from binary pulsar dragging effects (Yunes and Pretorius \cite{Yunes:2009hc} obtain $\beta\gtrsim 10^{-15}$ km$^{-4}$ from the pulsar PSR J0737-3039 A/B). This bound has been derived under the assumption that the BH is non-rotating, but it should provide a reliable estimate
on the range of allowed $\beta$ in the general case.

Furthermore, we found some evidence that in the limit $\beta\rightarrow\infty$, there are no
non-trivial QNM solutions: the only allowed solutions in this limit seem to be the ordinary
Schwarzschild gravitational QNM, with a vanishing scalar field.

Much more remains to be done. The coupled system of equations (\ref{eqq}) and (\ref{eqpsi}) was not
solved in general in the present paper, nor were the exact limits of instability investigated. It
would be extremely interesting to study the spectra of this system, either by analytical
approximations or by numerical means. A promising approach would consist in doing wave scattering
using time-evolution methods \cite{Gundlach:1993tp}, and reading off the quasinormal modes directly
from the frequency and decay time of these perturbations.

The formalism developed here can be used to study point particles in the background geometry and to
compute the gravitational wave signal generated by extreme-mass-ratio inspirals. Generalization to
rotating black holes is a major (rotating black hole solutions are only partially
understood \cite{Alexander:2009tp}), but fundamental task to accomplish.
\section*{Acknowledgements}
We thank Nico Yunes and Antonello Polosa for useful suggestions and discussions. This work was partially supported by FCT - Portugal through projects PTDC/FIS/64175/2006, PTDC/FIS/098025/2008 and PTDC/FIS/098032/2008.

\appendix

\section{Recurrence relations}\label{recurrence}
Here we try to generalize the continued fraction method \cite{Leaver:1985ax,Nollert:1993zz} to solve
Eqns (\ref{eqq}), (\ref{eqpsi}).

The functions $Q,\Theta$ with the correct asymptotic behavior can be written as
\begin{eqnarray}
Q&=&\left(\frac{r}{2M}-1\right)^{-2\ii M\omega}
\left(\frac{r}{2M}\right)^{4\ii M\omega}
e^{\ii\omega(r-2M)}\Phi\nonumber\\
\Theta&=&\left(\frac{r}{2M}-1\right)^{-2\ii M\omega}
\left(\frac{r}{2M}\right)^{4\ii M\omega}
e^{\ii\omega(r-2M)}\Sigma\,,
\end{eqnarray}
where $\Psi,\Sigma$ can be expressed as
\begin{equation}
\Phi=\sum_{n=0}^\infty a_ny^n\,,~~~~~~\Sigma=\sum_{n=0}^\infty b_ny^n
\end{equation}
in terms of the dimensionless variable
\begin{equation}
y\equiv 1-\frac{2M}{r}\,.
\end{equation}
Note that $a_{n<0}=b_{n<0}=0$; we set the overall normalization by imposing $a_0=1$, and the (complex) constant $b_0$ is an unknown of the problem.

Replacing these expressions in the perturbation equations (\ref{eqq}), (\ref{eqpsi})
we find the recurrence relations
\begin{eqnarray}
&&a_n\alpha_n+a_{n-1}\beta_n+a_{n-2}\gamma_n+\lambda\left(
b_{n-2}\sigma_n^1+\dots+b_{n-7}\sigma_n^6\right)\nonumber\\
&&b_n\alpha_n+b_{n-1}\tilde\beta_n+b_{n-2}\tilde\gamma_n
+b_{n-3}\tilde\delta_n+\rho\left(
a_{n-2}\sigma_n^1+\dots+a_{n-7}\sigma_n^6\right)\nonumber\\\label{contfraceq}
\end{eqnarray}
where
\begin{eqnarray}
\alpha_n&=&n(n-4\ii M\omega)\nonumber\\
\beta_n&=&-2n^2+(16\ii M\omega+2)n-8\ii M\omega+32M^2\omega^2-l(l+1)-1
\nonumber\\
\gamma_n&=&n^2-(8\ii M\omega+2)n+8\ii M
\omega-16M^2\omega^2+1\nonumber\\
\tilde\beta_n&=&\beta_n+4+\frac{9\pi l(l+1)}{\beta M^4}\nonumber\\
\tilde\gamma_n&=&\gamma_n-4-\frac{54\pi l(l+1)}{\beta M^4}\nonumber\\
\tilde\delta_n&=&\frac{135\pi l(l+1)}{\beta M^4}\nonumber\\
\lambda&=&\frac{3\pi\ii\omega}{M^4}\nonumber\\
\rho&=&\frac{(l+2)!}{(l-2)!}
\frac{3\ii\omega}{16\beta M^4}\nonumber\\
\vec\sigma_n&=&(1,-5,10,-10,5,-1)\,.
\end{eqnarray}
The two complex recurrence relations (\ref{contfraceq}), depending on the two unknown complex quantities $(\omega,b_0)$, in principle converge only for a discrete set $(\omega,b_0)_j$, corresponding to the QNM; it is not clear how to numerically implement this condition, since it seems not possible to express the recurrence relations (\ref{contfraceq}) in terms of continued fractions.

\section{Asymptotic behavior of the perturbation equations}\label{app:asympt}
We first note that, as $r_*\rightarrow\pm\infty$ (i.e. $r\rightarrow
r_H\equiv2M,r\rightarrow+\infty$), the perturbation equations (\ref{eqq}) and (\ref{eqpsi}) reduce
to the simple coupled wave equations
\begin{eqnarray}
&&\left(\frac{d^2}{dr_*^2}+\omega^2\right)\Theta=-\frac{f}{r^5}\frac{(l+2)!}{(l-2)!}\frac{6iM}{\omega\beta}Q\,,\\
&&\left(\frac{d^2}{dr_*^2}+\omega^2\right)Q=-\frac{f}{r^5}96\pi iM\omega\Theta\,.
\end{eqnarray}
Changing scalar wavefunction 
\begin{equation}
\Theta^{\rm old}\to \frac{1}{4\omega} \sqrt{\frac{(l+2)!}{(l-2)!\pi\beta}} \,\Theta^{\rm new}
\end{equation}
we get
\begin{eqnarray}
&&\left(\frac{d^2}{dr_*^2}+\omega^2\right)\Theta^{\rm new}=-\frac{f\,T}{r^5}Q\,,\\
&&\left(\frac{d^2}{dr_*^2}+\omega^2\right)Q=-\frac{f\,T}{r^5}\Theta^{\rm new}\,,
\end{eqnarray}
with $T=\sqrt{\frac{(l+2)!}{(l-2)!}}24iM\sqrt{\pi/\beta}$. Finally, defining $\Psi^+\equiv
Q+\Theta^{\rm new}$ and $\Psi^-\equiv Q-\Theta^{\rm new}$, one gets the equivalent system
\begin{eqnarray}
&&\left(\frac{d^2}{dr_*^2}+\omega^2\right)\Psi^{+}=-\frac{f\,T}{r^5}\Psi^{+}\,,\\
&&\left(\frac{d^2}{dr_*^2}+\omega^2\right)\Psi^{-}=\frac{f\,T}{r^5}\Psi^{-}\,.
\end{eqnarray}
Each of the wavefunctions $\Psi^{\pm}$ has a simple harmonic behavior at the boundaries and
therefore (\ref{behaviorboundaries}) follows.

\end{document}